\documentclass{appolb}
\usepackage{epsfig}

\usepackage{amsmath}
\usepackage{amssymb}
\usepackage{graphicx}

\begin{document}
\preprint{MIT-CTP-4044}
\title{Anthropic constraints on fermion masses %
\thanks{Presented at ExcitedQCD 09, Zakopane, Poland.}%
}
\author{Alejandro Jenkins
\address{Center for Theoretical Physics, Laboratory for Nuclear Science and Department of Physics,
Massachusetts Institute of Technology, Cambridge, MA 02139, USA}
}
\maketitle
\begin{abstract}
We summarize the results of previous research on the constraints imposed on quark masses by the anthropically-motivated requirement that there exist stable nuclei with the right charge to form complex molecules.  We also mention an upper bound on the mass of the lightest lepton, derived from the requirement that such nuclei be stable against electron capture.
\end{abstract}
\PACS{12.38.Aw, 14.65.Bt, 21.10.Dr}
  
\section{Introduction}
\label{sec:intro}

It is conceivable that some of the properties of the physical world might not be uniquely determined by an underlying dynamical principle, but might instead reflect the requirement that those properties be compatible with the existence of intelligent observers like us, capable of studying them.  In modern theoretical physics, this controversial idea is called the ``{\it anthropic principle}'' \cite{carter}.

Broadly speaking, two camps are inclined to take the anthropic principle seriously, which we may call the ``best-of-worlds'' camp and the ``worst-of-worlds'' camp.  The former, notably represented by the authors of \cite{barrow}, holds that since the parameters of physical laws seem finely tuned to allow life, the universe is somehow geared ---perhaps by a superior intelligence--- towards being hospitable to us.  The second camp, represented by \cite{susskind}, holds that our universe is only {\it barely} capable of accommodating intelligent observers, which is what one might expect if the fundamental physical laws are not designed with us in mind, but allow for the existence of many universes with different properties, only a very few of which {\it happen} to be compatible with the evolution of intelligent life.

The German philosopher Arthur Schopenhauer captured this distinction almost two centuries ago:
\begin{quote}
Against the palpably sophistical proofs of Leibniz that this is the best of all possible worlds, we may even oppose seriously and honestly the proof that it is the worst of all possible worlds. For possible means not what we may picture in our imagination, but what can actually exist and last. Now this world is arranged as it had to be if it were to be capable of continuing with great difficulty to exist; if it were a little worse, it would be no longer capable of continuing to exist. Consequently, since a worse world could not continue to exist, it is absolutely impossible; and so this world itself is the worst of all possible worlds. \cite{schopenhauer}
\end{quote}
If we replace Schopenhauer's ``which can exist and last'' with ``{\it which can be observed},'' and his ``capable of continuing to exist'' with ``{\it capable of sustaining intelligent observers},'' his argument almost fits into the contemporary debate in theoretical physics.

In recent years, the anthropic principle ---in the worst-of-worlds sense--- has gained support among theoretical physicists primarily for two reasons: first, Weinberg used anthropic reasoning to predict a small, positive cosmological constant \cite{weinberg} before it was convincingly measured \cite{darkenergy}.  Weinberg's argument seems particularly compelling in light of the absence of any convincing dynamical explanation for the smallness of the cosmological constant.  Second, string theory, which is generally regarded as the most serious candidate for a theory of quantum gravity, is now believed to predict a vast landscape of possible vacua, which could lead to a multiverse ---populated by eternal inflation--- in which causally disconnected regions exhibit very different low-energy physics \cite{douglaskachru}.

Regardless of how seriously one takes the claim that the anthropic principle {\it explains} some feature of the physical laws we have observed, one can reasonably ask how the parameters of the Standard Model (SM) are {\it constrained} by the requirement that complex structures (such as are presumably required for intelligent life) be possible.  This is the point of view taken in \cite{congenial}, where we tried to quantify how far the masses of the three light quarks ($u$, $d$, and $s$) could be modified while preserving some stable form of the chemical elements hydrogen (charge 1) and carbon (charge 6), which are necessary to form complex molecules.  In this work, universes that have stable forms of both hydrogen and carbon are labeled {\it congenial} and those that lack one or both are labeled {\it uncongenial}.  Congenial worlds were generally found to have stable forms of oxygen (charge 8) and other heavy elements that seem to be important to life in our own universe.  This work has also been summarized in \cite{perez}.

\section{Slices and approximations}
\label{sec:slices}

In this sort of exercise, one must first define what ``slice'' through the parameter space of the SM is to be explored.  The parameters of the SM that are directly relevant to nuclear physics ---and therefore to the question of whether organic chemistry is possible--- are: the mass  $m_e$ of the electron, the scale $\Lambda_{\rm QCD}$ below which QCD becomes strongly coupled, the masses $m_q$ of the quarks which are light with respect to $\Lambda_{\rm QCD}$, and the electrical charges of the light quarks.\footnote{It will not be necessary to assume anything about the CKM matrix, other than that it should have no accidental zeroes.}

The results of \cite{congenial}, summarized in Sec. \ref{sec:quarks}, correspond to a slice along which $m_e$ is fixed to its value in our universe, while $\Lambda_{\rm QCD}$ is adjusted in order to keep the average mass of the lightest baryon flavor multiplet fixed to what it is in our world.  We only consider worlds with at most three light quarks, not due to any theoretical prejudice but simply because those are the only worlds for which we can reliably estimate baryon masses, by using first-order perturbation theory in the breaking of flavor $SU(3)$ symmetry, with the corresponding parameters extracted from the spectrum of baryon masses in our own world.

Dimensional analysis suggests that all scales $M_i$ relevant to nuclear physics ---including the baryon masses as well as the masses of the $f_0(600)$, $\rho(770)$ and $\omega(783)$ resonances that determine nuclear binding energies\footnote{The pion and kaon masses vary as the square root of the light quark masses, which is extremely steep close to the chiral symmetry point, but single pion and kaon exchanges are not important in determining nuclear binding energies: being Goldstone bosons, they are derivatively coupled and therefore do not contribute in the $s$-wave.}--- should vary roughly as
\begin{equation}
\frac{\Delta M_i}{M_i^\oplus} \sim \frac{\Delta m_q}{\Lambda_{\rm QCD}^\oplus}~,
\end{equation}
where $\Delta m_q$ is the light quark mass variation being considered, and the superscript $\oplus$ indicates the value of the parameter in our own world.  Since we consider $\Delta m_q$'s only up to order $\sim 10$ MeV, we expect $\Delta M_i$'s of only a few percent.

Nuclear structure is exquisitely sensitive to the baryons masses: for instance, the $\Lambda$ baryon could be as little as 20 MeV heavier than the proton and the neutron and still not form stable nuclei.  The reason for this is well understood, and can be seen even in a simple Fermi gas model of the nucleus \cite{congenial}.  We therefore keep the variation in the baryon masses (which we can compute using perturbation theory) and ignore the quark-mass dependence of the nuclear binding energies (which cannot be reliably computed at the moment).

It should also be pointed out that some of the universes that have stable forms of hydrogen and carbon might never efficiently synthesize those elements.  Previous research has explored the effects on nucleosynthesis of varying the light quark masses \cite{hogan}, but mapping out the history of universes different from our own and determining for which of them the elements required for organic chemistry never become sufficiently abundant to sustain life anywhere, seems too ambitious a task, given our current level of understanding.

\section{Three light quarks}
\label{sec:quarks}

\begin{figure} [b]
\begin{center}
\includegraphics[width=0.5\textwidth]{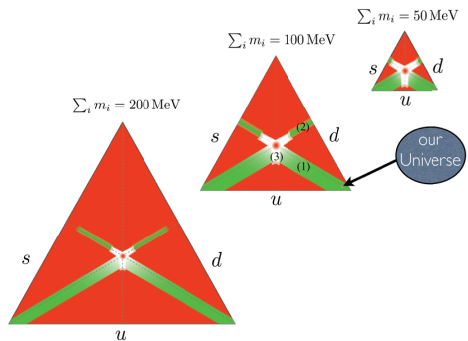}
\end{center}
\caption{\small Graphical representation of the space of three light quark masses, taken from \cite{perez}.  The mass of one of the light quarks is given by the distance to the corresponding side of the triangle, with the altitude of a given triangle corresponding to a fixed value of $m_u + m_d + m_s$.  Congenial worlds are green, and uncongenial worlds are red.  White areas cannot be definitively classified at this time.}
\label{fig:triangles}
\end{figure}

We estimate nuclear masses as functions of the baryon masses using either analog nuclei \cite{analog} or a generalized semi-empirical mass formula.  We can then check for the stability of nuclei against various strong and weak decay processes.  This allows us to categorize worlds as congenial or uncongenial, as shown in Fig. \ref{fig:triangles}.

The region marked (1) in Fig. \ref{fig:triangles} correspond to worlds like ours, with nuclei made of one positively charged and one neutral baryon.  The region marked (2) corresponds to worlds made of a neutral and a negatively charged baryon.  The region marked (3) (called the ``baryonic zoo'' in \cite{perez}) corresponds to worlds in which more than two species of baryons can form stable nuclei.  It is difficult to characterize these worlds in detail, but worlds in which flavor $SU(3)$ is nearly unbroken are very likely uncongenial, because heavy stable nuclei would tend to be electrically neutral.  This implies that there probably is a lower anthropic bound on $m_s$: If the mass of the $s$ quark were smaller than its observed value in our world by more than about an order of magnitude, it is unlikely that a stable form of carbon would exist.

\section{Lightest lepton}
\label{sec:leptons}

Only the lightest lepton will be stable against weak decays, and the next lepton will not have a long enough lifetime to be relevant to chemistry unless there is a very near degeneracy in mass.  For simplicity, we shall refer in general to the lightest lepton as the ``electron.''

Clearly, there must be an anthropic upper bound on the electron mass, when other physical parameters are held fixed.  In particular, the Bohr radius must be greater than the size of the nucleus for long-lived atoms to exist at all.  Since the Bohr radius $a$ is inversely proportional to $m_e$, and since in our world $a = 0.5 \times 10^{-10}$ m, while nuclei have a size of order $\sim 10^{-15}$ m, worlds in which the electron is more than $\sim 10^5$ times heavier than it is in our world are anthropically forbidden.

A more restrictive constraint on the electron mass comes from avoiding too large a spontaneous rate for the weak fusion process $p p \rightarrow \,^{2}\mbox{H} \, e^+ \, \nu_e$.  As is well known from the study of muon-catalyzed fusion \cite{jackson}, a heavier electron would enhance the rate of fusion due to the greater overlap of the wavefunctions of the protons in a hydrogen molecule.  A back-of-the-envelope calculation suggests that if the electron were more than about two orders of magnitude heavier than it is, fusion would prevent a mole of water from ever cooling below its boiling point.

The best constraint on the electron mass, however, seems to come from avoiding the instability of nuclei to decays mediated by the capture of an atomic electron.  If we keep track of the dependence on $m_e$ in the analysis of \cite{congenial}, we see that there will be no stable form of hydrogen unless
\begin{equation}
m_e < M_n - M_p + B(^{3}\mbox{H}) = M_n - M_p + 8.5 ~\hbox{MeV}~,
\end{equation}
where $B(^{3}\mbox{H})$ is the binding energy of the triton.  A bound of the same order is obtained from requiring that heavy nuclei not be unstable to weak neutron emission\footnote{This is a process, analogous the decay of a hypernucleus in our world, in which the nucleus emits a neutron, via a weak interaction.}
\begin{equation}
m_e \lesssim \frac{1}{2} \left(M_n - M_p \right) + \frac{dB_{\rm max}(A)}{d A}
\end{equation}
where $dB_{\rm max} / d A$ is the binding energy per nucleon for nuclei at to bottom of the valley of stability, which for medium to heavy nuclei $\approx 8$ MeV.  Therefore, the stability of heavy nuclei also requires $m_e \lesssim 10$ MeV.

Whether there might also be a {\it lower} anthropic bound on the electron mass is a more difficult question.  The congeniality bounds of Sec. \ref{sec:quarks} for worlds where $M_n > M_p$ came from stability of carbon isotopes against $\beta$ decay, but these would change little even if the electron were massless. Lowering the electron mass would decrease the energy of ionization of hydrogen and other elements, which would change the temperature of recombination and alter the late history of the universe.  It would also change the temperature scale of organic processes.  It is, however, difficult to translate these considerations into sharp statements about which universes would contain intelligent life and which would not.

{\bf Acknowledgements}:  The author thanks Allan Adams and Bob Jaffe for discussions leading to the results of Sec. \ref{sec:leptons}, and Gilad Perez for permission to use Fig. \ref{fig:triangles}.  This work was supported in part by the U.S. Department of Energy under contract DE-FG03-92ER40701.


\bibliographystyle{aipprocl}   


\end{document}